\documentclass[aps,prl,twocolumn,showpacs,groupedaddress]{revtex4}

\usepackage[dvips]{graphics}
\begin{document}

\title{Why is the condensed phase of DNA preferred at higher temperature?\\
DNA compaction in the presence of a multivalent cation}

\author{Takuya Saito}
%\email[]{saito@chem.scphys.kyoto-u.ac.jp}
\affiliation{Department of Physics, Graduate School of Science, Kyoto University, Kyoto, 606-8502, Japan}

\author{Takafumi Iwaki\footnote{Present address: Department of Chemistry, College of Staten Island, The City University of New York 10314, USA}}
%\email[]{iwaki@chem.scphys.kyoto-u.ac.jp}
\affiliation{Department of Physics, Graduate School of Science, Kyoto University, Kyoto, 606-8502, Japan}
%%\altaffiliation{Present address: Department of Chemistry, College of Staten Island, The City University of New York 10314, USA}

\author{Kenichi Yoshikawa\footnote{To whom correspondence should be addressed. Tel:+81-75-753-3812. Fax:+81-75-753-3779. Email:yoshikaw@scphys.kyoto-u.ac.jp}}
%\email[To whom correspondence should be addressed. Tel:+81-75-753-3812. Fax:+81-75-753-3779. Email:]{yoshikaw@scphys.kyoto-u.ac.jp}
\affiliation{Department of Physics, Graduate School of Science, Kyoto University, Kyoto, 606-8502, Japan}

\date{\today}

\begin{abstract}

Upon the addition of multivalent cations, a giant DNA chain exhibits a large discrete transition from an elongated coil into a folded compact state. We performed single-chain observation of long DNAs in the presence of a tetravalent cation (spermine), at various temperatures and monovalent salt concentrations. We confirmed that the compact state is preferred at higher temperatures and at lower monovalent salt concentrations. This result is interpreted in terms of an increase in the net translational entropy of small ions due to ionic exchange between higher and lower valence ions.

\end{abstract}


\maketitle

\def\degC{\kern-.2em\r{}\kern-.3em C}

The environment presented by a solution can dramatically affect the morphology of polyelectrolyte molecules. For example, upon the addition of multivalent cations such as spermine(4+) or spermidine(3+), a giant DNA molecule undergoes a large discrete transition from an elongated coil into a folded compact state accompanied by a change in volume on the order of $10^{-4}$~-fold. Most previous experimental and theoretical investigations reported that this so-called coil-globule transition should be continuous~\cite{Widom,Wilson_Bloomfield}. About a decade ago, based on the direct observation of individual DNA molecules by fluorescent microscopy, it has been clarified that DNA compaction is largely discrete, i.e., a first-order phase transition~\cite{YoshikawaPRL}. In the transition between the coil and compact states, the coil can be regarded as a disordered, disperse state~\cite{yoshinaga_afm}. In addition, it is known that the compact state shows an ordered morphology, such as a toroid~\cite{Gosule_toroid}. A recent preliminary study with single-molecule observation suggested that the compact state is preferred at higher temperature~\cite{murayama}, despite the significant decrease in conformational entropy in the compact state, although there remained the possibility of significant influence due to kinetic effect. The purpose of the present study was to gain better insight into this unexpected effect of temperature on DNA compaction, through a careful experiment to discriminate thermodynamic from kinetic effects. Since DNA is a highly charged polyelectrolyte, the effects of counterions may be critically important in solving this problem. In this article, we report the dependence of DNA compaction on temperature and the salt concentration in the presence of spermine (a tetravalent cation) by means of single-chain observation, under careful considerations to attain thermal equilibrium state by avoiding the kinetic effect.

The observation of single DNA molecules was performed as described previously~\cite{YoshikawaPRL,murayama}, using fluorescence microscopy. Briefly, the experimental conditions were as follows. Bacteriophage T4DNA (166~kbp, Nippon Gene) was dissolved at 0.2~$\mu$M (in base units) in a 10~mM Tris-HCl buffer solution (pH~7.5) with 0.1~$\mu$M of the fluorescent dye 4'6-diamidino-2-phenylindole (DAPI, Wako Chemical Industries). We define the total monovalent salt concentration as the sum of the Tris-HCl~(10~mM) and NaCl~(Nacalai Tesque) concentrations. The temperature of the sample was controlled by a water-circulating system and a thermoplate~(Tokai hit). In the experiment on the effect of temperature, we prepared all of the solutions at 4~\degC{}, increased the temperature, and then observed the samples after 2 hours of incubation.

Figure~1(a) shows fluorescence images of a single T4DNA molecule; (left) elongated coil state with spermine at 0.2~$\mu$M and (right) folded compact state with spermine at 3.0~$\mu$M. To characterize the size of DNA, we measured the long-axis length, $L$, which is defined as the longest distance in the outline of DNA images in Figure~1(c). Figure~1(d) shows the dependence of $L$ of T4DNA molecules on the concentration of spermine, at a fixed monovalent cation (Tris-HCl) concentration (10~mM) and temperature (21~\degC{}). At 0.2~$\mu$M spermine, all of the DNA macromolecules show an elongated coil conformation with a mean $L$ value of ca.~3~$\mu$m. On the other hand, when the spermine concentration is higher than 2.0~$\mu$M, all of the DNA chains show a folded compact state. At intermediate concentrations, the elongated coil and compact states coexist. Thus, DNA compaction is a discrete transition (first-order transition) at the level of individual DNA molecules~\cite{YoshikawaPRL}.

\begin{figure}
\includegraphics{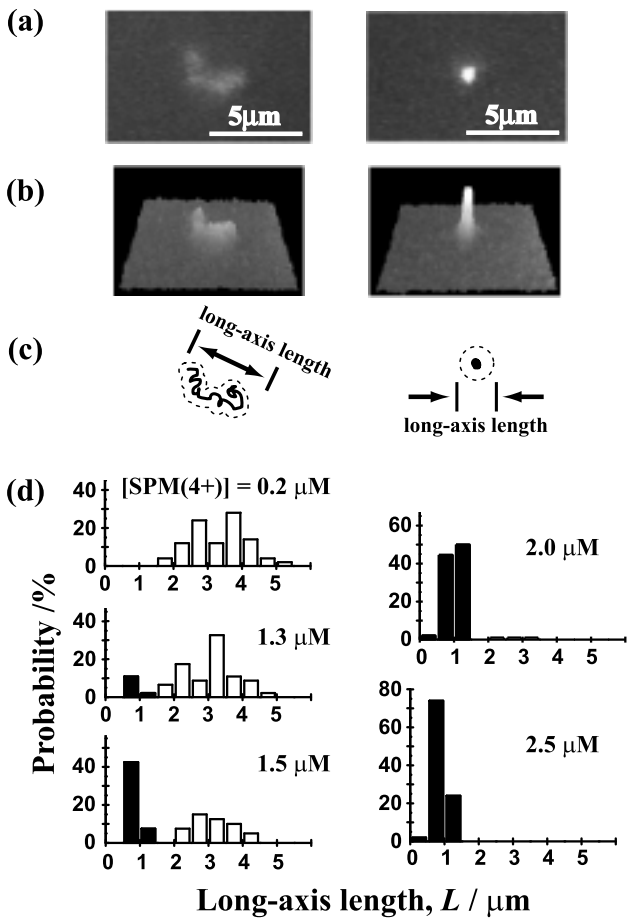}
\caption{(a) Fluorescence images of single T4DNA molecules: (left) elongated coil state at 0.2 $\mu$M spermine; (right) folded compact state at 3.0 $\mu$M spermine. (b) Quasi-3D representation of fluorescent light intensity, corresponding to the fluorescence image. (c) Schematic representations of the fluorescence images and the actual conformations. (d) Histogram of the long-axis length, $L$, of T4DNA molecules as it varies with the spermine concentration, at a fixed monovalent cation concentration~(10~mM) and temperature (21~\degC{}). }
\label{fig1}
\end{figure}

\begin{figure}
\includegraphics{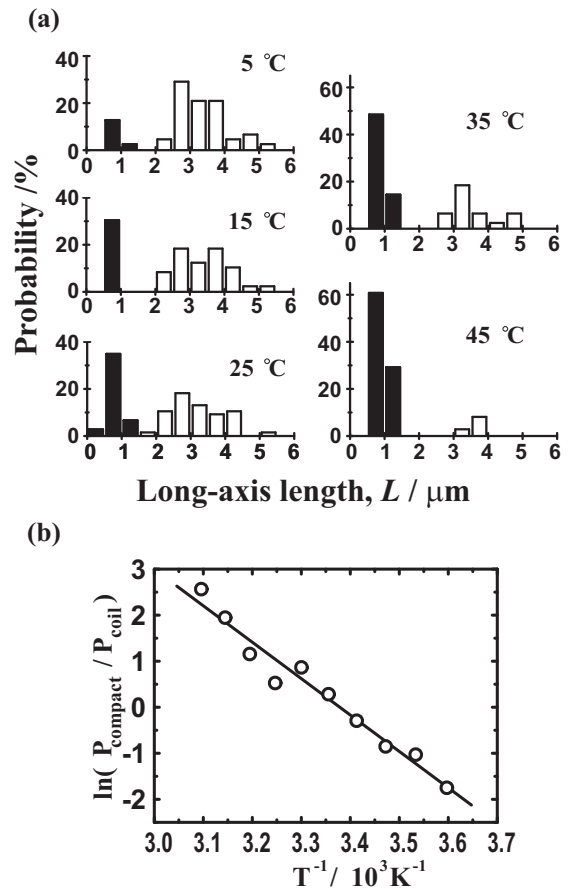}
\caption{Histograms of the long-axis length, $L$, at different temperatures. The filled and open bars indicate the compact and coil states, respectively. (b) Temperature-dependence of the relative population of the coil and compact states.}
\label{fig2}
\end{figure}

We next examined the effect of temperature. Figure~2 shows the effect of temperature on the size distribution of DNA molecules at a fixed concentration of spermine (1.5~$\mu$M) and monovalent cation (10~mM). The population of the compact state increases with an increase in temperature, indicating that the condensed phase is preferred at higher temperatures. In Figure~2(b), the logarithm of the relative population of compact and coil states is plotted as a function of $1/T=k_B\beta$ (van't Hoff Plot). The plot is interpreted using the equation
\begin{eqnarray}
\log{(P_{\rm{coil}}/P_{\rm{compact}})}=-\beta \Delta G=-\beta \Delta H +\Delta S /k_B
\end{eqnarray}
Thus, the enthalpy difference is estimated to be $\Delta H \approx 30k_B T_c$, where $T_c$ is the transition temperature for this solution. At the transition point, $\Delta G$ is equal to zero. The entropy change from the coil to the compact state is therefore 
\begin{eqnarray}
\Delta S (=S_{\rm{compact}}-S_{\rm{coil}})\approx +30 k_B
\end{eqnarray}
Next, we examined the effect of the monovalent salt concentration. Figure~3 shows a diagram of the DNA conformation as a function of the concentrations of a monovalent cation and tetravalent cation (spermine) at a fixed temperature (21~\degC{}). The gray zone indicates the region of coexistence of the coil and compact states. The critical spermine concentration increases with the monovalent cation concentration and DNA compaction is suppressed with an increase in the monovalent salt concentration. 

In Figure~2, the compact state is preferred at higher temperatures, indicating that the system entropy increases with DNA compaction. The appearance of a condensed state at a higher temperature is contrary to the expected result regarding temperature dependence~\cite{Toyoichi_Styrene_temperature}, since it is generally understood that an elongated coil has greater entropy than the compact state due to elastic entropy. Let us discuss this positive change in entropy by considering the thermodynamics of individual chains. For the sake of simplicity, we consider the following form of free energy for a single DNA chain
\begin{eqnarray}
F_{\rm{total}}(\alpha,\theta_i)=F_{\rm{ela}}(\alpha)+F_{\rm{ele-st}}(\alpha,\theta_i)+F_{\rm{trans}}(\theta_i)
\end{eqnarray}
where ~$F_{\rm{total}}(\alpha,\theta_i)$ is the total free energy, $F_{\rm{ela}}(\alpha)$ originates from the elastic entropy of a single DNA chain,~$F_{\rm{ele-st}}(\alpha,\theta_i)$ is the electrostatic energy,~$F_{\rm{trans}}(\theta_i)$ is the translational entropy of small ions, $\alpha$ is a swelling parameter of DNA, and $\theta_i$ is the ratio of condensed ion around the macro molecule and subscript $i$ represents small ions (monovalent cation, tetravalent cation and monovalent anion).

\begin{figure}
\includegraphics{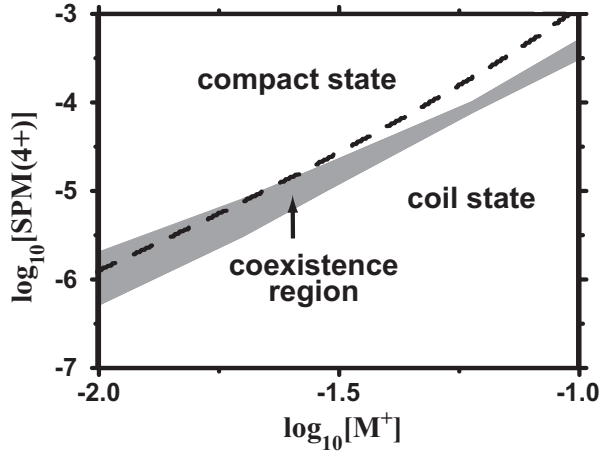}
\caption{Phase diagram as a function of the monovalent and tetravalent cation concentrations at 21~\degC{}. The gray zone indicates the region of coexistence of the coil and compact states. The broken line is the calculated line deduced from the change in translational entropy of $\Delta S_{\rm{trans}}=+1000k_B$ with $\lambda =50$,~$\eta =1.5$.}
\label{fig3}
\end{figure}

The elastic free energy can be written~\cite{RedBook} as
\begin{eqnarray}
\beta F_{\rm{ela}}(\alpha)=\frac{3}{2}(\alpha^2+\alpha^{-2})
\end{eqnarray}
By substituting the hydrodynamic radius of the coil and compact states~\cite{N-G2}, we obtain $\Delta S_{\rm{ela}} \approx -1000k_B$.

The translational entropy of small ions (tetravalent cations, monovalent cations and monovalent anions) can be written as
\begin{eqnarray}
\beta F_{\rm{trans}}(\theta_i)=Q \sum_{i=4+,1+,1-}\theta_i \log{\frac{\theta_i}{e c_i v_i}} 
\end{eqnarray}
where $c_i$ is the small ion concentration in the bulk, $Q~(=3.32\times 10^3 e)$ is the total charge number of a T4DNA chain, $v_i$ is the effective volume ($v_{1+}=v_{1-}=v_{4+}/4=8\pi(\xi-1)d^3$), $d$ is the charge spacing in the DNA chain, $\xi(=l_{\rm{B}}/d)$ is the Manning parameter, and $l_{\rm{B}}$ is the Bjerrum length.

Based on the 2-phase nature of a DNA chain, we consider independent reduced formulations for $F_{\rm{ele-st}}$ for each state. The electrostatic free energy for the coil state is
\begin{eqnarray}
\beta F_{\rm{ele-st}}^{\rm{coil}}(\alpha^{\rm{coil}},\theta_i^{\rm{coil}})&=& \nonumber \\
&-&Q \xi(1-\sum_{i} \theta_i^{\rm{coil}})^2\log({1-\exp^{-\kappa d}}) \nonumber \\
&+&Q \lambda(\theta_{4+}^{\rm{coil}}-\eta \theta_{1-}^{\rm{coil}})^2
\label{coil_ele}
\end{eqnarray}
where $ \kappa^{-1}$ is the Debye length, and $\lambda$ and $\eta$ are the interaction parameters between the condensed tetravalent cations and monovalent anions. In eq.(\ref{coil_ele}), the first term is derived from Oosawa-Maning condensation theory\cite{Oosawabook,Manning_QRB_1978}, which explains the distribution of small ions around a polyelectrolyte chain in a good solvent. Near the transition point, as in the present case, the correlation between the small ions should be more significant. Thus, we introduce the second term which reflects the interaction between condensed multivalent cations and monovalent anions, since the condensed multivalent cations should attract monovalent anions to form a layer of polyanions, multivalent cations and monovalent anions. This should essentially be similar to the overcharge phenomenon~\cite{overcharge_Nguyen,overcharge_Tanaka}. The minimization of total free energy for the coil state gives us the equilibrium value of each ion ratio, $\theta_i^{\rm{coil}}$.

For simplicity, we do not discuss the explicit formulation of the electrostatic free energy for the compact state. Instead, the compact state is modeled as $4\theta_{4+}^{\rm{compact}} =1$ and $\theta_{1+}=\theta_{1-}=0$. We assume that the counterion inside the compact state can move around the segments even if the segments are fixed\cite{jmb_spermine_invisible}, and the compact state is neutralized electrically \cite{yamasaki_foldedstate}.

From the above estimation of the total and elastic entropies, we evaluate $\Delta S_{\rm{trans}}$ as follows.
\begin{eqnarray}
\Delta S_{\rm{trans}} = \Delta S -\Delta S_{\rm{ela}} \approx +1000k_B
\label{trans1000}
\end{eqnarray}
The dashed line in Figure~3 is determined from this equation~(\ref{trans1000}). This equation originates from the equation $\Delta G = \Delta H-T_c \Delta S = 0$, which means that the transition line occurs at $T_c$. This theoretical line behaves as an increasing function of the monovalent salt concentration. From the above analysis, the variation in the ion distribution by the binding of 1 tetravalent cation and sequent release of ions to the bulk can be estimated as $\sim3.0$ for monovalent cations and $\sim1.8$ for monovalent anions. This variation in the ion distribution can be regarded as ionic exchange between higher and lower valence ions of each sign. The higher valence ions, tetravalent cations and polyvalent anion (DNA chain), are assembled or condensed into compact DNA. On the other hand, lower valence ions, monovalent cations and monovalent anions, are repelled into the bulk solution. Thus, the positive change in entropy originates from the change in translational entropy upon such ion exchange. Thus, the ion distribution in the coil state contributes more to the decrease in the entropy of small ions than that in the compact state. Ionic exchange should be responsible for the temperature-dependence of DNA compaction. Our model explains the effects of both temperature and the monovalent salt concentration in compaction.

Some nonionic hydrophobic polymers (e.g. poly(N-isopropylacrylamide)~\cite{Heskins_NIPAAM,Hirotsu_NIPAAM}) are also known to undergo condensation at a higher temperature in water. In this case, temperature-dependence is explained by the effect of hydration. Since DNA is a highly charged polymer, it behaves as a hydrophilic polymer when its double-stranded structure is preserved. Thus, the temperature effect of hydration on such a hydrophobic polymer should be different from that with DNA. Actually, in pure water, DNA does not fold to a compact state at any temperature. This suggests that the hydration effect is not the main force for compaction of a DNA chain. Furthermore, hydrophobic interaction cannot explain the monovalent salt effect seen in the present experiment.

Recently, the appearance of the compact state in polyelectrolytes has attracted much interest from physicists in relation to the problem of "like-charge attraction". The importance of the correlation between the counterions has been stressed~\cite{Oosawabook,like-charge-attraction1,like-charge-attraction2,B.I.Shklovskii,Gelbart_two_rod,Rouzina_Bloomfield,Stevens,Leikin_zipper}. If the correlation between the ions is the driving force of the stabilization in the compact state, it would be reasonable to expect that the compact state should be preferred at lower temperatures. In contrast, however, our experiment showed an opposite trend. Many current theoretical studies concerning "like-charge attraction" fail to adequately describe this temperature-dependence, since these theories have not properly accounted for the change in translational entropy. The entropy of ion distribution in the ordered compact state is not necessarily less than that in the disordered coil state, even if multivalent cations are condensed around the polyelectrolytes to give "like-charge attraction". In addition, the dielectric constant of the solvent (water) roughly depends on temperature as $\epsilon \propto T^{-1.4}$~\cite{lange_handbook_chemistry}, and the Bjerrum length is then subject to $l_B \propto T^{0.4}$, which suggests that the electrostatic energy has a greater effect at higher temperature. It is possible that such a dielectric effect contributes to the apparent temperature-dependence of DNA compaction. However, at present, it is difficult to precisely evaluate the contribution of the dielectric effect in the transition. Nevertheless, clearly the effect of ionic exchange on the transition must be considered to interpret the effects of both temperature and the monovalent salt concentration.

In this study, we examined the effects of temperature and the monovalent salt concentration on the compaction of single DNA molecules. Our results suggest that ionic exchange plays an important role in the manner of transition, i.e., the effects of both temperature and the monovalent salt concentration on compaction originate from the increase in net translational entropy due to ionic exchange between higher and lower valence ions.

\bibliography{sci.bib}

\end{document}